# Development of interatomic potential appropriate for simulation of dislocation migration in fcc Fe

Mikhail I. Mendelev[1] and Valery Borovikov[1]

[1]*Division of Materials Sciences and Engineering, Ames Laboratory, Ames, IA 50011*

**Abstract**

Molecular dynamics (MD) simulation of dislocation migration requires semi-empirical potentials of the interatomic interaction. While there are many reliable semi-empirical potentials for the bcc Fe, the number of the available potentials for the fcc is very limited. In the present study we tested three EAM potentials for the fcc Fe (ABCH97 [Phil. Mag. A, **75**, 713-732 (1997)], BCT13 [MSMSE **21**, 085004 (2013)] and ZFS18 [J. Comp. Chem. **39**, 2420-2431 (2018)]) from literature. It was found that the ABCH97 potential does not provide that the fcc phase is the most stable at any temperature. On the other hand, the fcc phase is always more stable than the bcc phase for the BCT13, ZFS18 potentials. The hcp phase is the most stable phase for the BCT13 potential at any temperature. In order to fix these problems we developed two new EAM potentials (MB1 and MB2). The fcc phase is still more stable than the bcc phase for the MB1 potential but the MB2 potential provides that the bcc phase is the most stable phase from the upper fcc-bcc transformation temperature, $T_{\gamma \to \delta}$, to the melting temperature, $T_m$, and the fcc phase is the most stable phase below $T_{\gamma \to \delta}$. This potential also leads to an excellent agreement with the experimental data on the fcc elastic constants and reasonable stacking fault energy which makes it the best potential for the simulation of the dislocation migration in the fcc Fe among all semi-empirical potentials considered in the present study. The MD simulation demonstrated that only the ZFS18, MB1 and MB2 potentials are actually suitable for the simulation of the dislocation migration in the fcc Fe. They lead to the same orders of magnitude for the dislocation velocities and all of them show that the edge dislocation is faster than the screw dislocation. However, the actual values of the dislocation velocities do depend on the employed semi-empirical potential.

Steels are the most popular industrial materials because of their low cost and good mechanical properties. As a consequence, even an incremental improvement of their properties can have a profound economical effect and, therefore, these alloys were intensively studied for more than hundred years. Computer simulation approaches developed in the recent decades allow to reveal the atomic mechanisms of the processes which govern the alloy behavior under different conditions and therefore, guide further improvement of these materials. There are two main approaches in the atomistic simulation. In the first approach, an atomic mechanism is assumed and the energies associated with this mechanism are calculated at $T=0$. Then these energies can be used either in the transition state theory calculations or kinetic Monte Carlo (kMC) simulation to model the process at a finite temperature. The main advantage of this approach is the ability to use *ab initio* calculations at $T=0$ which is the most physically based approach to determine the interatomic interaction. But since one should couple the $T=0$ calculations with an assumed atomistic mechanism treated by means of a theory or in kMC simulation the resulting description is never really "*ab initio*". The problem becomes even more complicated if a high temperature application is the subject of a study since many theories incorporate the thermal contributions using the harmonic approximation. The classical molecular dynamics (MD) simulation offers an alternative approach where no approximations are made once a semi-empirical potentials of the interatomic interaction is defined. However, the results of a classical MD simulation heavily depend on the quality of the employed semi-empirical potential. Moreover, this potential should be computationally cheap enough to enable simulation of the atomic process under consideration. In studies of mechanical properties, obvious phenomena to consider are the dislocation migration and reactions. For a long time, the main obstacle preventing application of the MD simulation to study these processes in austenitic steels has been the absence of reliable semi-empirical potentials.

Since the main component of steels is Fe, a semi-empirical potential should be able to describe the main properties of this metal. At zero pressure, pure Fe exists in two allotropic configurations: body-centered cubic (bcc) phase which is stable from $T=0$ to $T_{\alpha\rightarrow\gamma}=1183$ K and from $T_{\gamma\rightarrow\delta}=1667$ K to the melting temperature, $T_m$, and face-centered cubic (fcc) phase which is stable from $T_{\alpha\rightarrow\gamma}$ to $T_{\gamma\rightarrow\delta}$. Since semi-empirical potentials are mostly fitted to $T=0$ properties, there are several types of semi-empirical potentials designed to simulate the bcc Fe. For example, a pair potential was proposed in [1], embedded atom method (EAM) [2] and Finnis-Sinclair (FS) [3] potentials were proposed in [4-6], modified embedded atom method (MEAM) [7] potentials were proposed in [8,9], a bond-order potential (BOP) [10-12] was proposed in [10]. A special effort was made to account for the magnetism of the bcc Fe at low temperatures (e.g., see [13-15]) which is out of the scope of the present study. The number of semi-empirical potentials suitable for the simulation of the fcc Fe is much smaller. Some of the potentials mentioned above are not even capable to provide mechanical stability of the fcc Fe (e.g., the FS potential developed in [5]). This obviously makes it difficult to use them as the starting point for development of semi-empirical potentials for austenitic steels.

The computational efficiency of a semi-empirical potential is another concern, which should be taken into account. The hard sphere model is obviously the most computationally efficient approach, but it is not suitable for the reproduction of the mechanical properties of metals. The pair interaction model is the next simplest interatomic approximation, but this model necessarily leads to zero Cauchy pressure (the difference between $C_{12}$ and $C_{44}$ elastic constants for the fcc/bcc metals) and, therefore, is also not suitable for the description of the mechanical properties of metals. EAM/FS potentials provide the same order of computational efficiency as pair potentials but capable to reproduce experimental values of the Cauchy pressure. This is why this type of potential is a very attractive choice to simulate the mechanical properties of steels. In the present study, we tested three EAM potentials proposed for the fcc Fe in the literature focusing on their performance at high temperatures, where the fcc is the most stable phase. We will refer them using the first letters of their author's names and the year the potential was published. The first potential is ABCH97 [4] which was developed primary for the bcc Fe. However, this potential does provide a metastable fcc phase and was recently used in the simulation of the fcc-bcc phase transformation [16,17]. The other two potentials, BCT13 [18] and ZFS18 [19], were chosen because they were specially developed to simulate the Fe-Ni-Cr austenitic steels. Since we were not quite satisfied by the performance of these EAM potentials we developed two new EAM potentials for the simulation of the fcc Fe at high temperatures as will be described below. It should be emphasized that having in mind simulations of the dislocation migration in the fcc Fe we developed the new potentials namely for this phase and did not try to describe the bcc Fe phase at low temperatures.

The rest of the manuscript is organized as follows. First, we will present the results of the testing of the EAM potentials taken from the literature. Next, we will present the new EAM potentials developed within the present study. Finally, we will present the results of the MD simulation of the dislocation migration in the fcc Fe obtained using all of the potentials mentioned above.

## 2. Testing of existing EAM potentials for fcc Fe

Any semi-empirical potential is an approximation and cannot describe all materials properties. Therefore, it is important to select the materials properties which are critical for a particular application. In this section we list the properties, acceptable description of which is required for a reasonable MD simulation of the dislocation migration and investigate how well the ABCH97, BCT13 and ZFS18 potentials satisfy these requirements.

The minimal requirements a potential should satisfy to be employed in a simulation of the dislocation migration in the fcc Fe are the following. First, the fcc phase should be stable enough, such that it never transforms into any other phase during MD simulation of the plastic deformation under experimentally realistic applied stresses. This requirement seems to be trivial because usually semi-empirical potentials are fitted to the cohesive energies of several competing crystal phases. However, if this is done at $T=0$ there is no warranty that the correct relations between the cohesive energies will hold at high temperatures. The obvious competing phase in the case of Fe is the bcc phase. Therefore, a potential which never leads to the formation of the bcc phase should be acceptable for the simulation of the plastic deformation in the fcc Fe. On the other hand, since in reality the difference between the free energies of these phases is not very large it is desirable (but not really necessary!) that a potential is able to reproduce this difference which is equivalent to reproduction of the bcc-fcc-bcc transformation temperatures and the corresponding latent heats and changes in the atomic volume.

Second, the semi-empirical potential should provide the correct elastic constants and staking fault energies (SFE). Third, if the dislocation climb is under consideration, the potential should provide the correct vacancy formation and migration energies. Since the sum of these quantities is the activation energy for the self-diffusion, the comparison with the experimental data on self-diffusion should be an excellent test of the reliability of a semi-empirical potential.

The self-interstitial formation energy is usually considered to be important only for the simulation of the radiation damage because the self-diffusion in the fcc metals is controlled by the vacancy mechanism. However, the interatomic interaction at small atomic separations which are present in any interstitial configurations is similar to those in activation processes. Therefore, it is desirable to provide at least reasonable values for the self-interstitial formation energies.

Since usually the potential development procedures include fitting semi-empirical potentials to $T$=0 properties we start testing of the literature EAM potentials by validating them against these properties. Several benchmark quantities are presented in Table I. Clearly the ABCH97 which was actually fitted to these quantities (except of the differences in the phase energies) provides the best results. The BCT13 and ZFS18, which were developed to simulate the fcc phase, obviously provide much worse agreement with the target properties listed in Table I.

Now we turn to testing the literature EAM potentials by the examination of the differences in the free energies, $\Delta G_{ss}$, between bcc, hexagonal close packed (hcp) and fcc phases at finite temperatures. In order to determine $\Delta G_{ss}$, we first determined the equilibrium lattice parameters and the latent heats associated with melting of these crystal phases as functions of temperature. Next, we determined the melting temperatures, $T_m$, of these phases using the protocol described in details in [20] which leads to the inaccuracy in $T_m$ about $0.005 T_m$. Then we calculated the changes in the free energies upon melting, $\Delta G_m$, via the Gibbs-Helmholtz equation. Finally, the differences between the free energies of solid phases were obtained as the differences between the corresponding values of $\Delta G_m$. The temperatures where $\Delta G_{ss}$=0 are the solid-solid transformation temperatures which can be determined this way with inaccuracy within 2 K as was discussed in [21]. Note that this approach does not rely on any approximation and its accuracy is only limited by the inaccuracy of the MD simulation.

The obtained differences in the free energies are shown in Fig. 1. This figure shows that the bcc phase is the most stable phase from 0 to $T_m$ for the ABCH97 potential, the hcp phase is the most stable phase from 0 to $T_m$ for the BCT13 potential and the fcc phase is the most stable phase from 0 to $T_m$ for the ZFS18 potential. This does not necessarily mean that the ABCH97 and BCT13 are not suitable for the simulation of the fcc Fe properties because MD simulation time is usually too short to observe the first order phase transformations but, obviously, one should be cautious employing these potentials. We will return to this issue in Section 4.

The SFEs as functions of temperature obtained from MD simulation are shown in Fig. 2. The SFEs obtained with the ABCH97 and BCT13 potentials are very small which is consistent with the very small differences in energies between the fcc and hcp phases (see Fig. 1). The SFE for the fcc Fe was obtained from *ab initio* calculations in [22]. It was found to be negative at low temperatures and positive above 300-500 K (the value depends of whether the magnetic entropy term was taken into account). At high temperatures the SFE should be around 50 mJ/m$^2$ according to the same calculations. The authors of [22] also pointed out that the extrapolation of the experimental data on the SFE in steels at room temperature obtained in [23] to the pure fcc Fe leads to 34 mJ/m$^2$ which reasonably agrees with their *ab initio* data. Since this is the free energy, the energy is supposed to be somewhat higher. Therefore, the ZFS18 potential leads to a reasonable SFE at least at high temperatures while the ABCH97 and BCT13 potentials dramatically underestimate it.

The elastic constants as functions of temperature obtained from the MD simulation are presented in Fig. 3. This figure shows that the temperature dependences of the elastic constants are not monotonic and different potentials lead to different results especially at low temperatures. On the other hand, all tested potentials lead to more similar results at high temperatures where the fcc phase is the ground state. The experimental data obtained at $T$=1428 K in [24] are also shown in Fig. 3. While the tested potentials provide an excellent agreement with the experimental value for $C_{11}$, the agreement for $C_{12}$ is much worse and the agreement for $C_{44}$ is not acceptable.

We now turn to the point defect properties. Table II shows the self-interstitial formation energies at $T$=0 in comparison with the *ab initio* values obtained in [25]. Only the ABCH97 potential provides reasonable magnitudes for these quantities while the BCT13 and ZFS18 potentials considerably overestimate the formation energies. The BCT13 potential even fails to predict that the <100> dumbbell is the most stable configuration. On the other hand, the ABCH97 and BCT13 potentials provide good agreement with the ab initio value of the vacancy formation energy at T=0 while the ZFS13 potential considerably overestimates it.

Since the fcc Fe is stable at high temperatures, it makes more sense to investigate predictions of a semi-empirical potential for the point defect properties at these temperatures rather than to make any conclusions based on the comparison with the *ab initio* calculations at $T$=0. In order to obtain the point defect formation and migration energies as functions of temperature we followed the method described in [26]. In this approach, one point defect (vacancy or self-interstitial) is introduced in a relatively small model (2048 atoms) and the energy is averaged over the course of the *NVT* (constant number of atoms, volume and temperature) MD simulation. Comparing this energy with that obtained from the perfect model at the same temperature and lattice parameter yields the point defect formation energy. The same MD simulation gives the effective atomic diffusivity in the models with point defects. Note that this diffusivity is different from the actual diffusivity because the point defect concentration in the simulation cell is not the equilibrium one. However, the activation energy of this effective diffusivity is the point defect migration energy. Finally, the sum of the point defect formation and migration energies is the activation energy for the self-diffusion in the particular mechanism.

The obtained results for the vacancy mechanism of the self-diffusion are shown in Fig. 4. This figure shows that the vacancy formation energy increases with increasing of temperature for all tested EAM potentials. However, the ABCH97 and BCT13 potentials lead to about the same vacancy formation energy while the ZFS18 potential leads to a much higher value. On the other hand the ZFS18 and BCT13 lead to about the same vacancy migration energy (1.1 eV/atom and 1.2 eV/atom, respectively; found from the slopes in the middle plot in Fig. 4) while the ABCH97 potential leads to somewhat higher value (1.6 eV/atom). All of these potentials predict very high activation energy for self-diffusion in comparison with the available experimental data, with the BCT13 potential leading to the smallest disagreement with experiment and the ZFS18 potential leading to the largest disagreement.

It is commonly accepted that the self-diffusion in the fcc metals proceeds via the vacancy mechanism (e.g., see [27]) because the self-interstitial formation energy is much higher than the vacancy formation energy such that the equilibrium vacancy concertation is much higher than the equilibrium self-interstitial concentration. The comparison of the data shown in Figs. 4 and 5 demonstrates that the self-interstitial formation energy is indeed much higher than the vacancy formation energy for all tested potentials at low temperatures. However, at high temperature these quantities become comparable. On the other hand, the self-interstitial migration energy is much smaller than the vacancy migration energy such that the activation energies for the self-diffusion in the vacancy and self-interstitial mechanism become comparable at very high temperatures. For example, at T=1600 K the BCT13 potential predicts 3.5 eV/atom for the vacancy mechanism and 3.2 eV/atom for the self-interstitial mechanism. Interestingly, a similar result was obtained for the bcc Fe in [28]. However, it was shown in [28], that it is still the vacancy mechanism which controls the self-diffusion at high temperatures, because the vacancy formation entropy at high temperatures is much higher than the self-interstitial formation entropy such that the equilibrium vacancy concentration is much higher than the equilibrium self-interstitial concentration.

To conclude, the ZFS18 potential is the only potential among the tested EAM potentials which provides that the fcc is the thermodynamically stable phase at high temperatures and this potential also leads to a reasonable SFE. However, the vacancy formation energy and the elastic constant $C_{44}$ obtained using this potential are much higher than the corresponding experimental values. Therefore, a more realistic potential is desirable.

### 3. New EAM potentials for the fcc Fe

We developed two new EAM potentials for the fcc Fe within the present study. The total potential energy in the EAM takes the following form:

$$U = \sum_{i=1}^{N-1} \sum_{j=i+1}^{N} \varphi(r_{ij}) + \sum_{i=1}^{N} \Phi(\rho_i) , \qquad (1)$$

where $N$ is the number of atoms in the system, $r_{ij}$ is the separation between atoms i and j, φ(r) is the pair potential, Φ(ρ) is the embedding energy function and

$$\rho_i = \sum_{\substack{j=1 \\ j \neq i}}^{N} \psi(r_{ij}) , \qquad (2)$$

where ψ(r) is the density function. This function is the same for both developed potentials while the pair potential and embedding energy functions are different.

Having in mind high temperature MD simulations of the dislocation migration in the fcc fcc phase where these potentials will be employed, we fitted the potential functions to the experimental elastic constants obtained at $T$=1428 K in [24] and an estimate of the SFE obtained in [22]. The cohesive energy and point defect formation energies at $T$=0 taken from [25] were also included in the potential development procedure with rather low weight. Since our previous potential for the bcc Fe [5] provided reasonable description of the solid-liquid interface properties of Fe (e.g., see [29]) we used this potential to generate liquid pair correlation function (PCF) at $T_m$ which was used as the target function for the development of the new potentials (the method to include a PCF in the potential development procedure was described in [30]).

The bcc phase was ignored in the development of the first potential (referred below as MB1). This is why it provides very poor agreement with the bcc phase properties at $T=0$ listed in Table I. Figure 1 shows that the bcc phase is not the most stable phase at any temperature. In the case of the fcc phase we choose the target value for $T_m$ to be a little below the actual melting temperature for Fe (1812 K) such that the fcc phase melts at 1795 K. Figure 2 shows that the MB1 potential somewhat overestimates the SFE but still provides a reasonable value. Figure 3 shows that the MB1 potential provides an excellent agreement with the experimental values for the elastic constants at $T=1428$ K. The examination of Table II shows that the MB1 potential provides rather reasonable agreement with the *ab initio* data on the self-interstitial formation energy but completely fails to reproduce the vacancy formation energy at $T=0$. The latter should not be viewed as a deficiency of the potential because it is not supposed to be employed in simulations at $T=0$. Figure 4 shows that with increasing temperature just above 20 K the vacancy formation energy quickly grows to a reasonable value and then shows the same trend as do the vacancy formation energies obtained with three semi-empirical potentials from the literature. Moreover, Fig. 4 shows that the MB1 potential predicts the activation energy for the self-diffusion in the fcc Fe in reasonable agreement with the available experimental data.

The main difference in the development procedure for the second potential (referred below as MB2) was that the bcc phase at high temperatures was considered. In particular, its melting temperature and bcc-fcc transformation temperature, $T_{\gamma \to \delta}$, were included in fitting. While the bcc phase is not even mechanically stable with the MB2 at $T=0$ (see Table I) Figure 1 demonstrates that the MB2 potential provides that the bcc phase is the most stable crystal phase at $T_m$. Its melting temperature is 1804 K which is just a little below the experimental value (1812 K). The fcc phase melts at T=1790 K and the bcc-fcc transformation temperature is 1702 K which is reasonably close to the experimental value (1667 K).

Figure 6 shows the atomic volume of the bcc and fcc phases as functions of temperature obtained using the MB2 potential. The bcc phase has a larger atomic volume below T=1666 K. In reality the bcc phase always has a larger atomic volume than the fcc phase in Fe. However, the experiment shows that the relative change in volume ($\Delta V/V = (V_\alpha - V_\gamma)/V_\alpha$) at $T_{\alpha \to \gamma}=1183$ K is 0.75 % while the MB2 potential leads to $\Delta V/V=0.86$ %. The relative change in volume drops to 0.5% when the temperature increases up to $T_{\gamma \to \delta}$. The MB2 potential leads to the same trend as in experiment but the relative change in volume decreases with increasing temperature 3 times faster. Therefore, this potential should be suitable for the MD simulation of the bcc-fcc transformation at $T=1250$ K where it leads to the same relative change in volume as in experiment. It can be also employed at other temperatures but one should be cautious about the effect of pressure on this transformation.

Inclusion of the bcc phase properties in the potential development procedure does not seem to worsen the reproduction of the materials properties critical for the simulation of the dislocation migration. Figure 2 shows that the MB2 potential provides a reasonable SFE and Fig. 3 shows that this potential also provides an excellent agreement with the experimental data on the fcc elastic constants. Figure 4 shows that the MB2 potential leads to somewhat higher vacancy formation energy comparing to that obtained using the MB1 potential. However, since the MB2 potential predicts a smaller vacancy migration energy comparing to that obtained using the MB1 potential (0.8 eV/atom vs. 1.1 eV/atom) both potentials lead to the same activation energy for the self-diffusion. Finally, Fig. 5 shows that both potentials lead to about the same values for the energies in the self-interstitial diffusion mechanism. Therefore, overall the MB2 potential looks like a better choice comparing to the MB1 potential. Yet, we used both potentials in the dislocation migration MD simulations described in the next section to see how the obtained results are sensitive to the choice of interatomic potential.

The developed potentials are available in the LAMMPS [31,32] format in [33,34]. The potential functions are shown in Fig. 7. They are pretty smooth which provides some warranty against unphysical behavior during MD simulation. The differences between the pair potential and embedding energy functions of these potentials are rather subtle such that it is difficult to explain looking at these functions why the MB2 potential better describes the Fe thermodynamic properties compared with those obtained using the MB1 potential. Finally, Fig. 7 shows the effective potential which is defined as

$$\varphi^{eff}(r) = \varphi(r) + 2 \frac{\partial \Phi}{\partial \rho}\bigg|_{\bar{\rho}} \psi(r) , \qquad (3)$$

where $\bar{\rho}$ is the average density determined via Eq. (2). These functions also look almost identical for both developed potentials.

## 4. Molecular dynamics simulation of the dislocation migration

All EAM potentials described in Sections 2 and 3 were employed in MD simulation of the pure edge and screw dislocation migration in the fcc Fe. The MD simulations were performed using LAMMPS [31,32] and the visualization was made using the OVITO software [35]. The snapshot images shown in Figs. 8-10 were obtained using the atomic positions averaged over 1 ps (to get rid of the thermal fluctuations). The simulation cells had periodic boundary conditions in all directions and contained a single dislocation dipole (edge or screw, as shown in Fig. 8). The simulation cell size was ~ 30×10×30 nm$^3$ s and it contained ~800,000 atoms. The details of the dislocation dipoles construction can be found in [36].

After constructing the edge/screw dipole configuration, the simulation cells were equilibrated at a desired temperature and zero applied stress for 60 ps. The MD time step was 2 fs. The parameter of the Nosé-Hoover thermostat determining how rapidly the temperature is relaxed was equal to 0.2 ps. The parameter of the barostat, determining how rapidly the stress is relaxed, was equal to 2 ps. Examples of snapshots obtained at $T$=1273 K are shown in Fig. 9. All dislocations shown in this figure dissociated into partial dislocations. The separation between them depends on the SFE value. Therefore, it is not surprising that the ABCH97 and BCT13 potentials lead to very large separations between partials because these potentials provide very low SFEs (see Fig. 2). The ZFS18, MB1 and MB2 potential lead to much smaller separations between partials. Similar results were obtained for the simulation cells containing screw dislocation dipoles (see Fig. 10).

A constant shear stress ($\sigma_{xz}/\sigma_{yz}$) was applied (in case of the edge/screw dipole configuration) to provide the driving force for the dislocation motion in the xy plane, in the x direction. This shear stress was gradually ramped up from zero to the target value during the time interval of 20 ps. The rest of the stress components were kept zero using the barostat. After reaching the target stress value, the system was kept under this constant shear stress additional 20 ps and then the dependence of the dislocation position on the simulation time was recorded. To find the dislocation positions at different simulation times we used the dislocation analysis modifier implemented in OVITO [37]. Since the dislocation is not necessarily straight in the y-direction, the average (over dislocation line x coordinates at different y coordinates) position in the x-direction was used. The dislocation velocities were determined as the slope of the dislocation position vs time lines. The obtained results are shown in Fig. 11. It should be noted that because of relatively small simulation cell sizes the effect of one dislocation on the motion of another can be substantial. This is probably why we did not observe any dislocation motion at small applied stresses. One should not confuse this effect with the Peierls stress. We will discuss this topic elsewhere [38] and in the present study we focus on the dependence of dislocation velocity on the employed semi-empirical potential.

All described above potentials were employed to study the dislocation migration at $T$=1273 K. It should be noted that we were able to obtain the data for the ABCH97 potential only for stresses not larger than 0.2 GPa (at higher stresses, the model completely transformed into the bcc lattice). We did not observe this phenomenon using any other tested potentials such that this is probably an artifact of the ABCH97 potential associated with the fact that this is the only potential providing that the bcc phase is the most stable phase at the simulation temperature. Also we could not obtain any data for the screw dislocation velocity using the BCT13 potential because once a stress was applied the separation between partials quickly became comparable with the simulation cell size.

Since namely the ZFS18, MB1 and MB2 potentials produce the most realistic dislocation behavior we will focus below only on these three potentials. If we ignore the lowest stress data, the dependence of the dislocation velocity on the applied stress can be divided into two regions: low velocity region where the dislocation velocity is proportional to the stress and high velocity region where the velocity rather weakly depends on the stress. At very small stresses (not shown in Figs. 11, 13), the velocity quickly drops to zero which is associated with the interaction between the dislocations of the dipole. The most encouraging result is that in spite of the differences in the potential development procedures used to develop the tested semi-empirical potentials they all lead to about the same order of magnitude of the dislocation velocity in both regimes. However, the difference in the velocities obtained using different semi-empirical potentials at moderate stresses (~0.4 GPa) can be two times and the difference at small stresses can be even larger.

Since the interaction between the dislocations in our simulation scheme does not allow to extract the dislocation mobility in the limit of small stress we investigated the temperature dependence of the dislocation velocities at a particular value of the stress (0.2 GPa). This choice was associated with the fact that this stress seems to be in the low velocity region for all considered potentials and temperatures (see Fig. 11). The obtained velocities as functions of temperature are shown in Fig. 12. In all cases the velocity decreases with increasing temperature in agreement with classical considerations (e.g., see [39]). This can explain why the dislocation velocities are higher for the ZFS18 potential than for the MB1 and MB2 potential. The fcc melting temperatures for the ZFS18, MB1 and MB2 potentials are 2417 K, 1795 K and 1790 K, respectively. If one assumes that the dislocation velocity actually depends on the homologous temperature, $T/T_m$, then at any given temperature, it is almost the same for the MB1 and MB2 potentials but much smaller for the ZFS18 potential. Therefore, the dislocation velocity for the ZFS18 potential should be higher than those for the MB1 and MB2 potentials. Figure 11 clearly supports this point for the case of the edge dislocation. Note that that the MB1 and MB2 potentials leads to almost the same values for the edge dislocation velocities. However, in the case of the screw dislocation they lead to different velocities. The screw dislocation velocity is the largest in the case of the ZFS18 potential, followed by the value obtained using the MB2 potential and the smallest value was obtained using the MB1 potential. Exactly the same relationship holds for the $C_{44}$ values (see Fig. 3) which may explain the obtained results.

Figure 13 shows the ratios of the edge and screw dislocation velocities under applied stress of 0.2 GPa. The edge dislocation is always faster (except one data point for the ZFS18 potential) however there is rather larger difference between considered potentials: this ratio is pretty close to 1 for the ZFS18 potential, somewhat larger for the MB2 potential and much larger for the MB1 potential. It is not clear which property of the considered potentials is responsible for such a difference.

## 5. Conclusions

In the present work, we tested three EAM potentials (ABCH97, BCT13 and ZFS18) from literature and developed two new EAM potentials (MB1 and MB2) suitable for the simulation of the dislocation migration in the fcc Fe. The ABCH97 potential does not provide that the fcc phase is the most stable at any temperature. On the other hand, the fcc phase is always more stable than the bcc phase for the BCT13, ZFS18 and MB1 potentials. The hcp phase is the most stable phase for the BCT13 potential at any temperature. The MB2 potential provides that the bcc phase is the most stable phase from $T_{\gamma \to \delta}$ to $T_m$ and the fcc phase is the most stable phase below $T_{\gamma \to \delta}$. Taking into account that this potential provides an excellent agreement with the experimental data on the fcc elastic constants and reasonable SFE, we conclude that the MB2 potential is the best potential for the simulation of the dislocation migration in the fcc Fe among all semi-empirical potentials considered in the present study. At the same time we emphasize that the MB1/MB2 potentials were specifically developed to simulate the fcc phase at high temperature. They should not be employed for the simulation of any Fe phase at low temperatures.

The MD simulation showed that only the ZFS18, MB1 and MB2 potentials are actually suitable for the simulation of the dislocation migration in the fcc Fe. They lead to the same orders of magnitude for the dislocation velocities and all of them shows that the edge dislocation is faster than the screw dislocation. However, the actual values of the dislocation velocities do depend on the employed semi-empirical potential.

**Acknowledgements:**

This work was supported by the National Energy Technology Laboratory (NETL) and United States Department of Energy's Office of Fossil Energy (USDOE-FE) Crosscutting Research Program, Briggs White, NETL Technology Manager, and Regis Conrad, USDOE-FE Program Manager. This work was conducted under the eXtremeMAT collaboration between Ames Laboratory, Idaho National Laboratory, Lawrence Livermore National Laboratory, Los Alamos National Laboratory, National Energy Technology Laboratory, Oak Ridge National Laboratory and Pacific Northwest National Laboratory; and executed through FWP AL-17-510091. Ames Laboratory is operated for the U.S. DOE by Iowa State University under contract # DE-AC02-07CH11358.

Table I. Physical properties calculated with EAM potentials at $T=0$.

| Property | Target value[§] | ABCH97[¥] | BCT13 | ZFS18 | MB1 | MB2 |
|---|---|---|---|---|---|---|
| a (Å), bcc | 2.855 | 2.867 | 2.860 | 2.867 | 2.884 | 2.817 |
| $E_{coh}$ (eV/atom), bcc | -4.316 | -4.316 | -4.280 | -4.280 | -4.248 | -3.977 |
| $C_{11}$ (GPa), bcc | 243 | 243 | 185 | 336 | 132 | -82 |
| $C_{12}$ (GPa), bcc | 145 | 145 | 154 | 171 | 113 | 112 |
| $C_{44}$ (GPa), bcc | 116 | 116 | 110 | 169 | 55 | 40 |
| a (Å), fcc | 3.658 | 3.680 | 3.562 | 3.633 | 3.549 | 3.548 |
| $\Delta E_{bcc \to fcc}$ (eV/atom) | 0.122 | 0.054 | -0.120 | 0 | -0.032 | -0.072 |
| a (Å), hcp |  | 2.602 | 2.519 | 2.570 | 2.498 | 2.496 |
| c/a, hcp |  | 1.633 | 1.633 | 1.593 | 1.676 | 1.687 |
| $\Delta E_{bcc \to hcp}$ (eV/atom) | 0.38 | 0.054 | -0.120 | 0.002 | -0.023 | -0.061 |

---

[§] The references for the target values can be found in [5].

[¥] This is the only potential in this table which was actually fit to reproduce $T=0$ properties.

Table II. Point defect formation energies (eV/atom) in the fcc Fe at $T=0$.

| Point defect | *ab initio* [25] | ABCH97 | BCT13 | ZFS18 | MB1 | MB2 |
|---|---|---|---|---|---|---|
| Vacancy | 1.67-1.95 | 1.94 | 1.87 | 2.78 | 0.29 | 1.05 |
| <100> dumbbell | 3.20-3.62 | 3.64 | 6.57 | 5.15 | 3.44 | 3.37 |
| <110> dumbbell | 4.17-4.82 | 4.08 | 5.71 | 5.51 | 3.99 | 3.82 |
| <111> dumbbell | 4.56 | 3.85 | 5.26 | 5.87 | 3.58 | 3.77 |

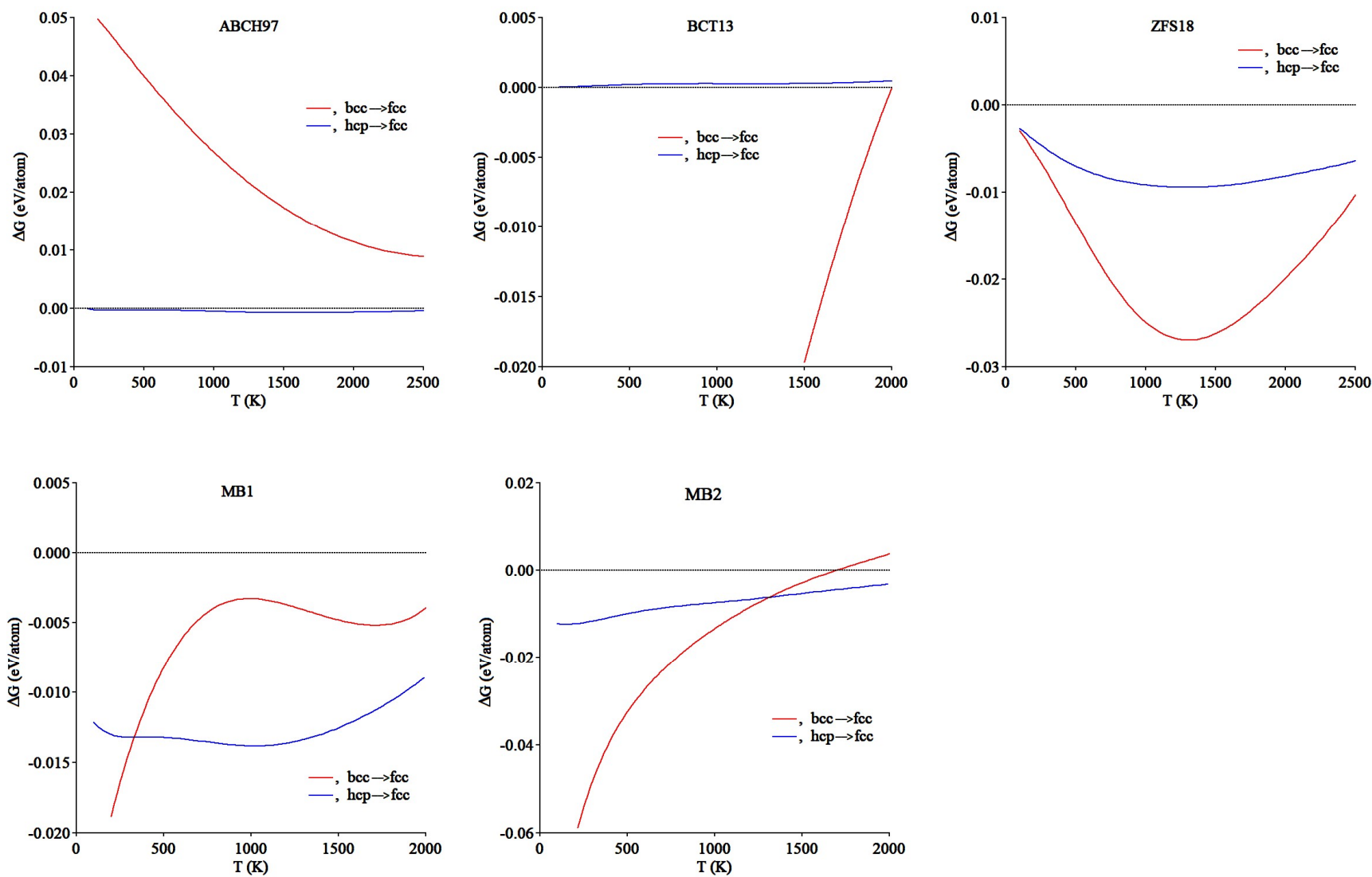

Figure 1. The difference between the free energies of competing crystal phases and the fcc phase. The negative value of this difference means that the fcc phase is more stable.

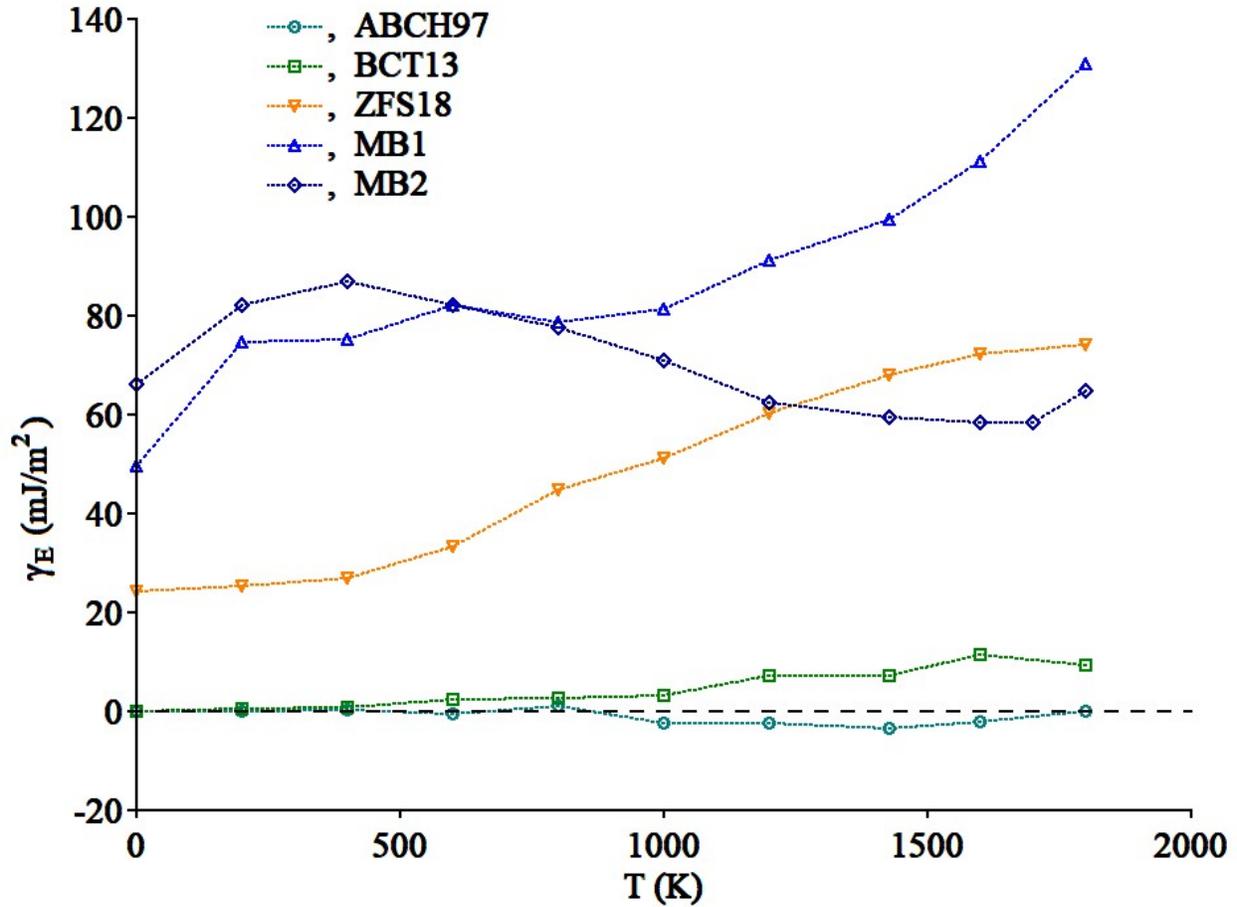

Figure 2. Stacking fault energy (not free energy!) as function of temperature. The staking fault free energy should be around 50 mJ/m$^2$ at high temperatures according to the *ab initio* calculations performed in [22]. The extrapolation of the experimental data on the staking fault free energy in steels at room temperature obtained in [23] to the pure fcc Fe leads to 34 mJ/m$^2$. Since this is the free energy, the energy is supposed to be somewhat higher.

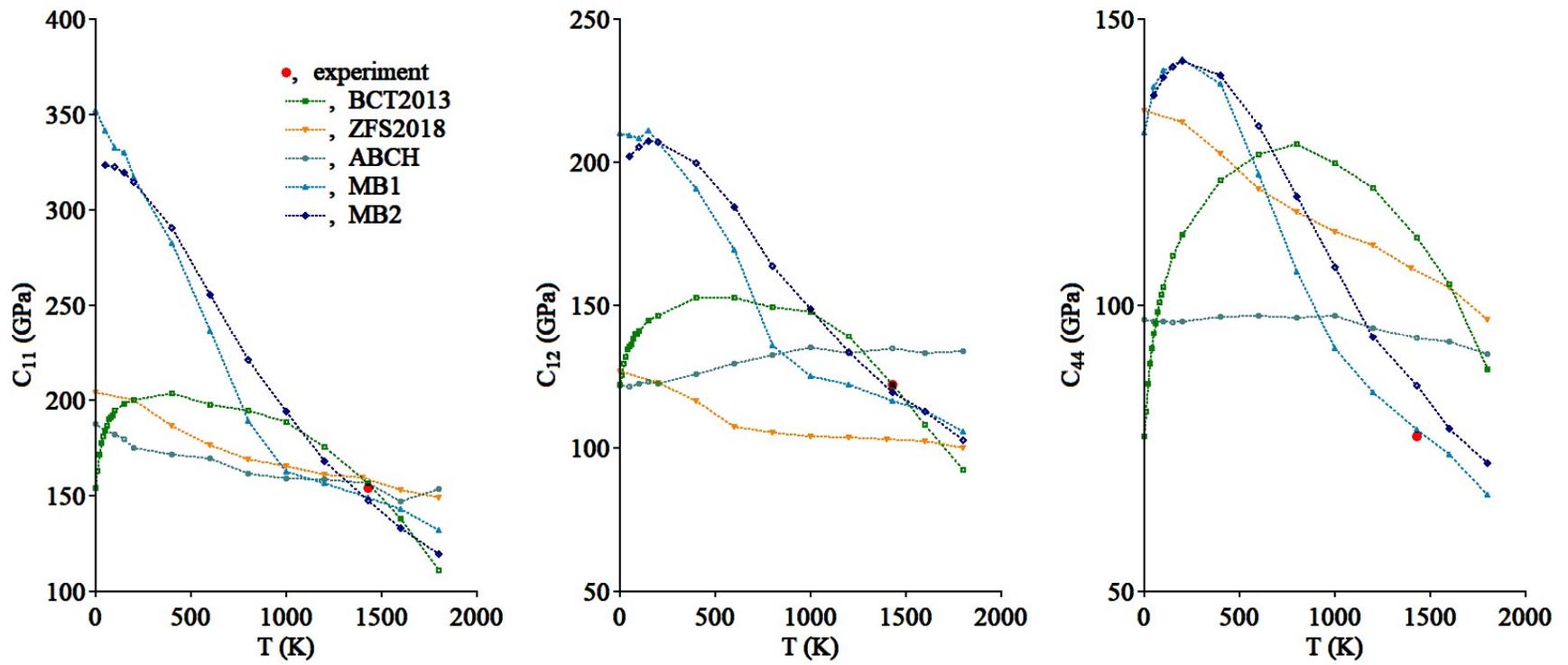

Figure 3. Elastic constants of the fcc Fe as functions of temperature.

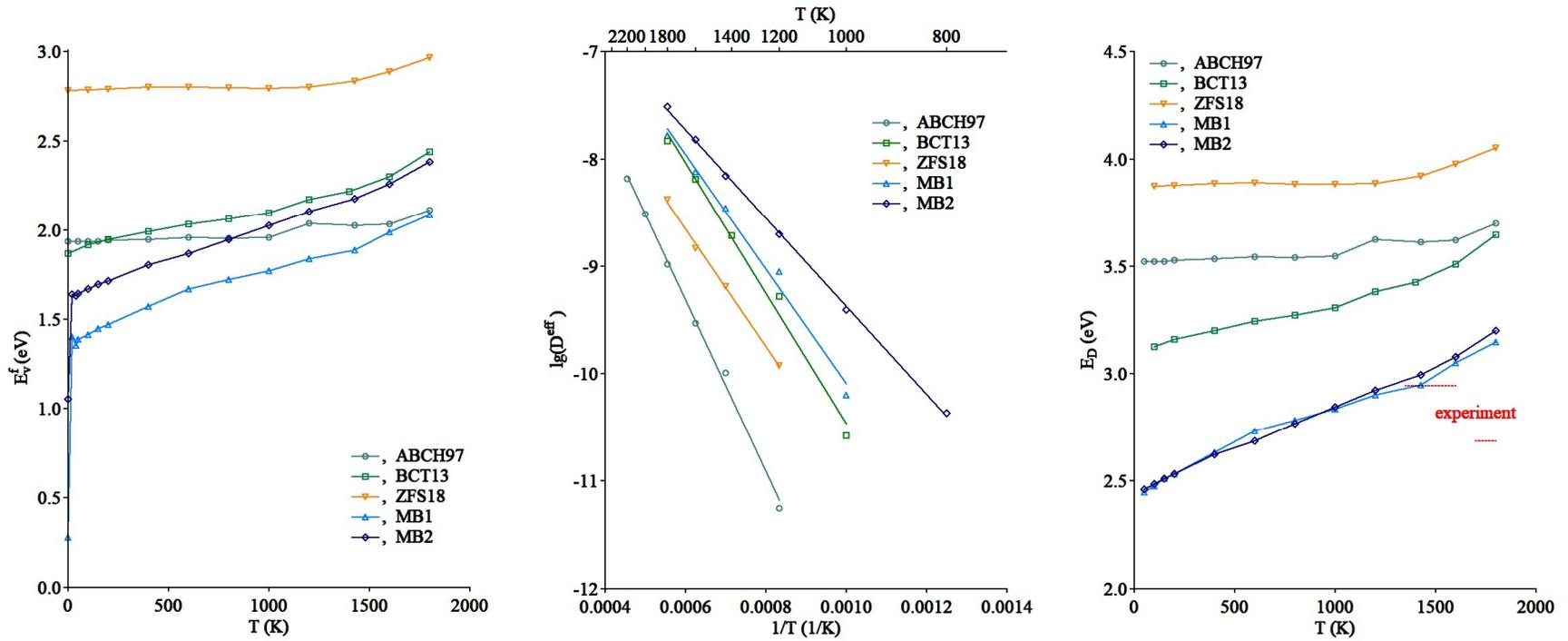

Figure 4. Parameters of self-diffusion in the vacancy mechanism: the vacancy formation energy; effective vacancy diffusivity and the activation energy for the self-diffusion (the details of calculations can be found in [26]). The red dashed lines correspond to two experimentally determined values of the activation energy of the self-diffusion in the fcc Fe taken from [40,41] (the widths of the lines correspond to the temperature intervals in which the data were obtained).

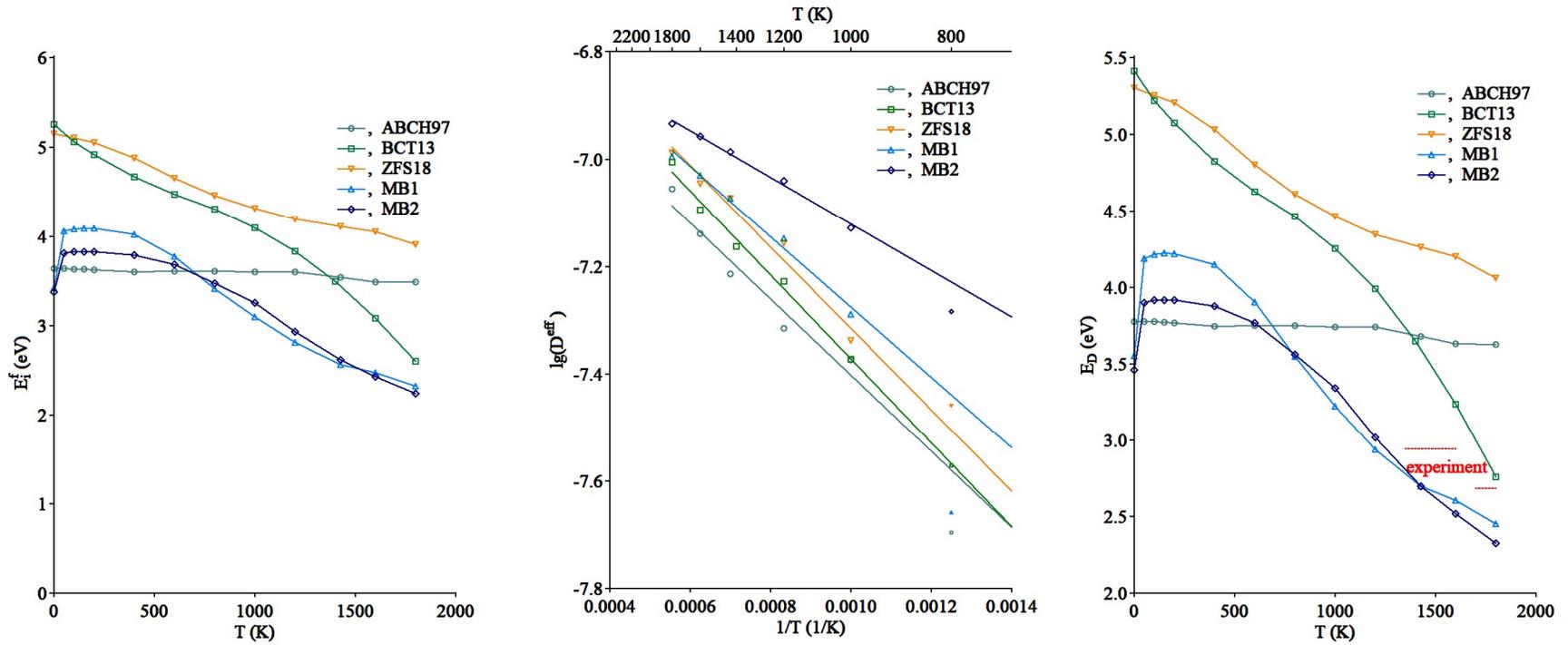

Figure 5. Parameters of self-diffusion in the interstitial mechanism: the vacancy formation energy; effective vacancy diffusivity and the activation energy for the self-diffusion (the details of calculations can be found in [26]). The red dashed lines correspond to two experimentally determined values of the activation energy of the self-diffusion in the fcc Fe taken from [40,41] (the widths of the lines correspond to the temperature intervals in which the data were obtained).

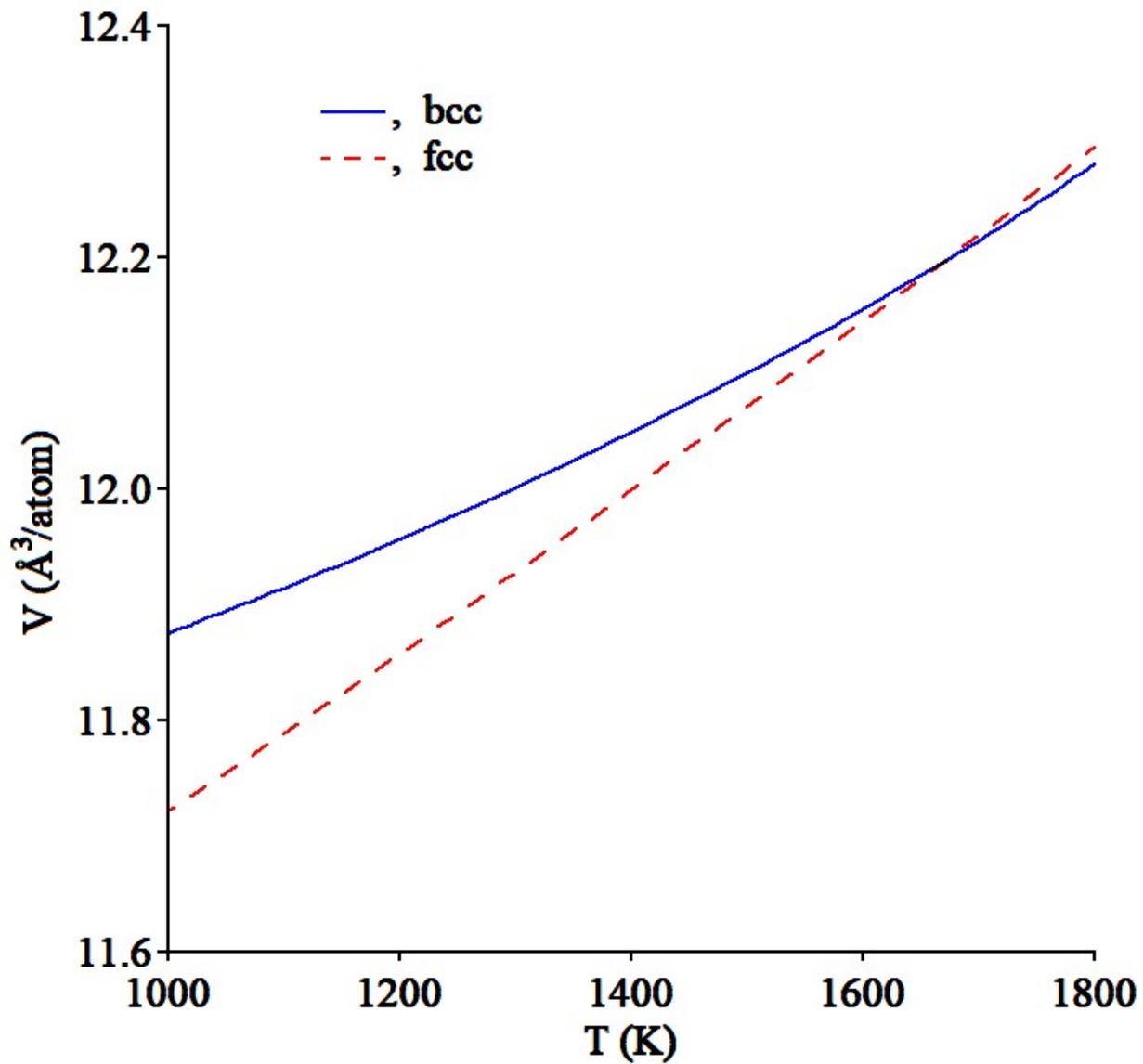

Figure 6. Atomic volumes of the bcc and fcc phases obtained using the MB2 potential.

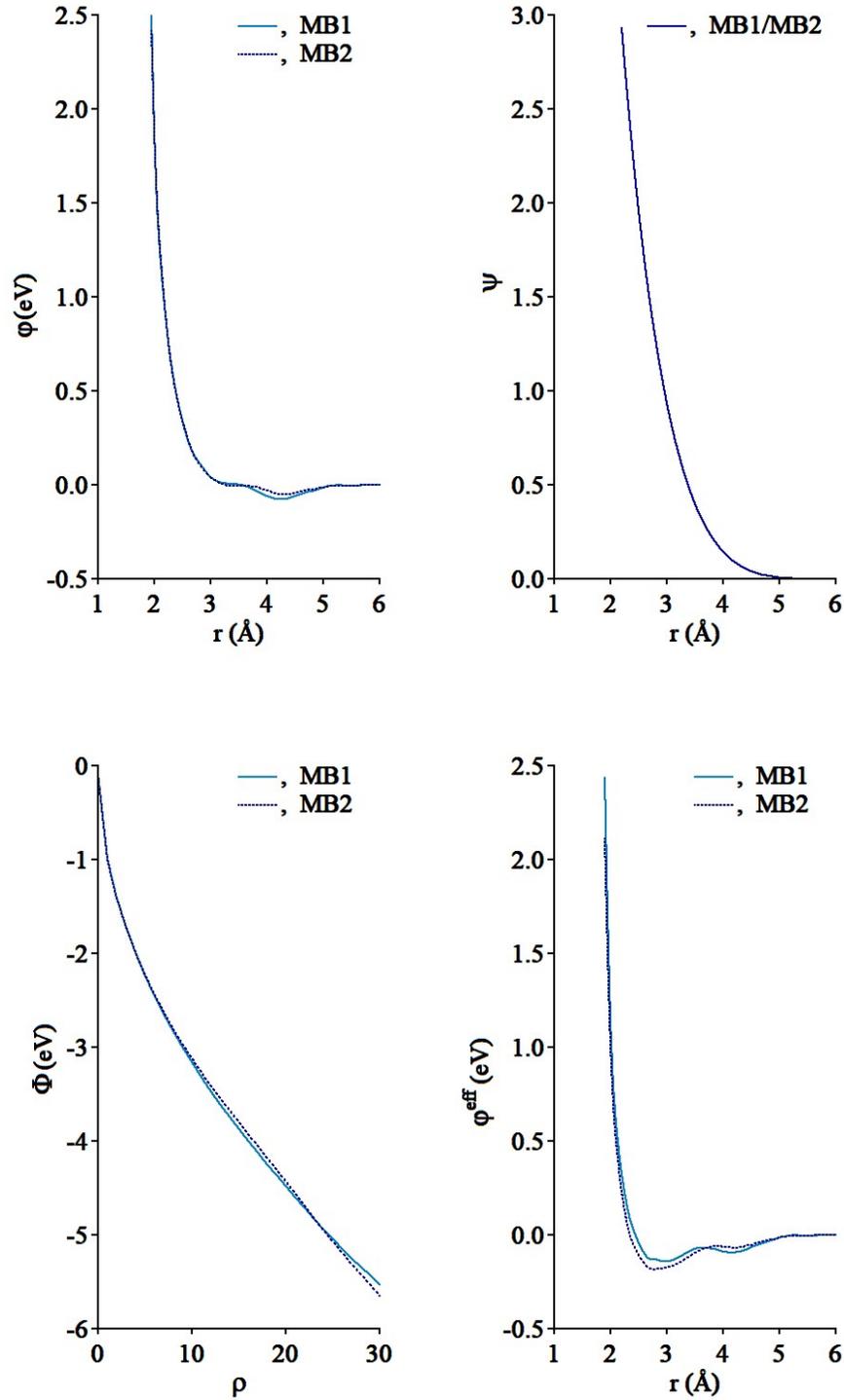

Figure 7. The developed potential functions. The effective potentials are calculated for the fcc phase. All functions are pretty smooth which provides some warranty against unphysical behavior during MD simulation.

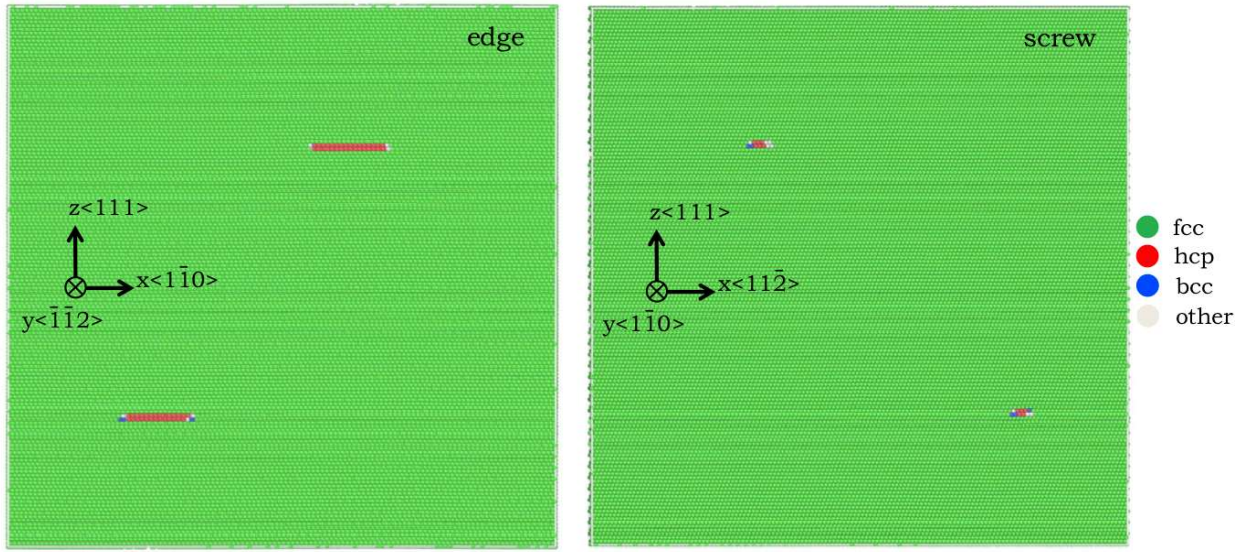

Figure 8. Simulation cells with edge and screw dislocation dipoles. The particular snapshots shown in this figure were generated using the MB2 potential at $T$=1273 K.

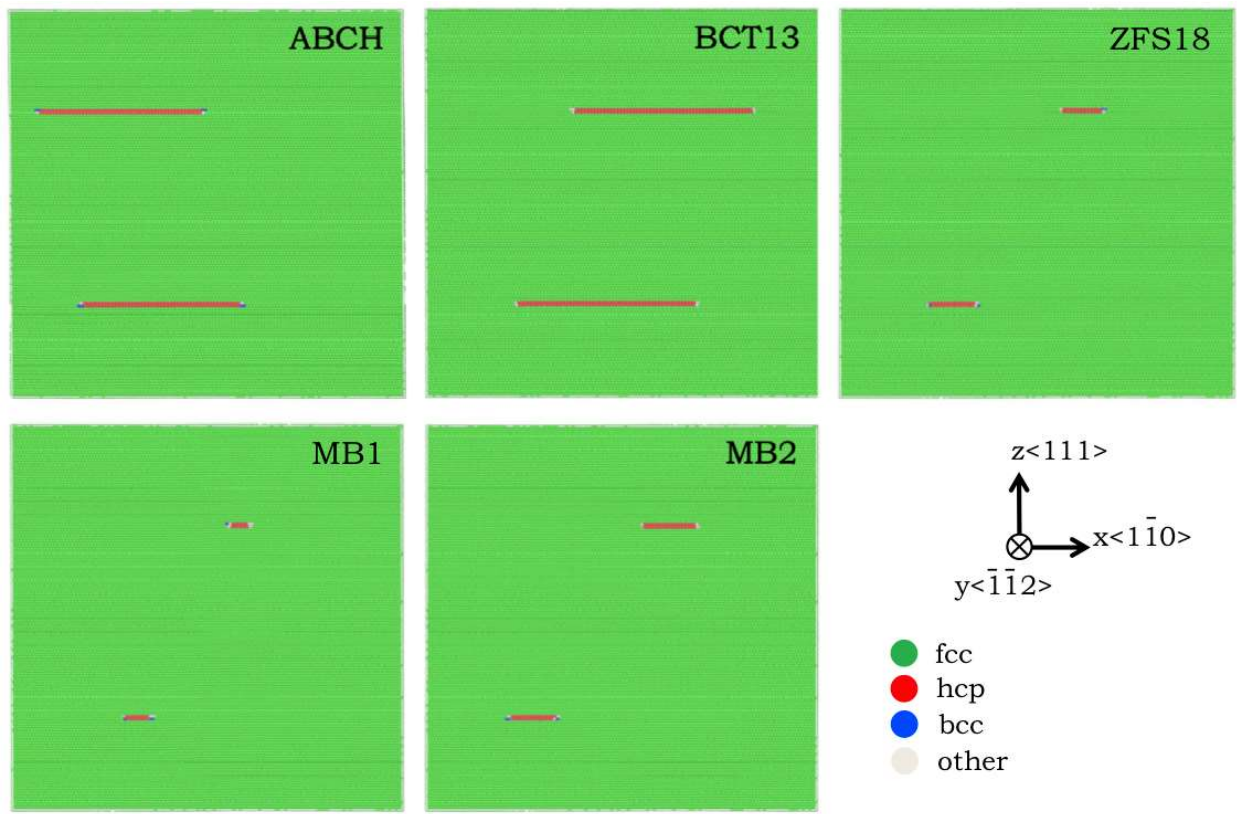

Figure 9. Snapshots of the simulation cells with edge dislocation dipoles at $T$=1273 K. The ABCH97 and BCT13 potentials lead to very large separations between partials because these potentials provide very low SFEs. The ZFS18, MB1 and MB2 potential lead to much smaller separations between partials.

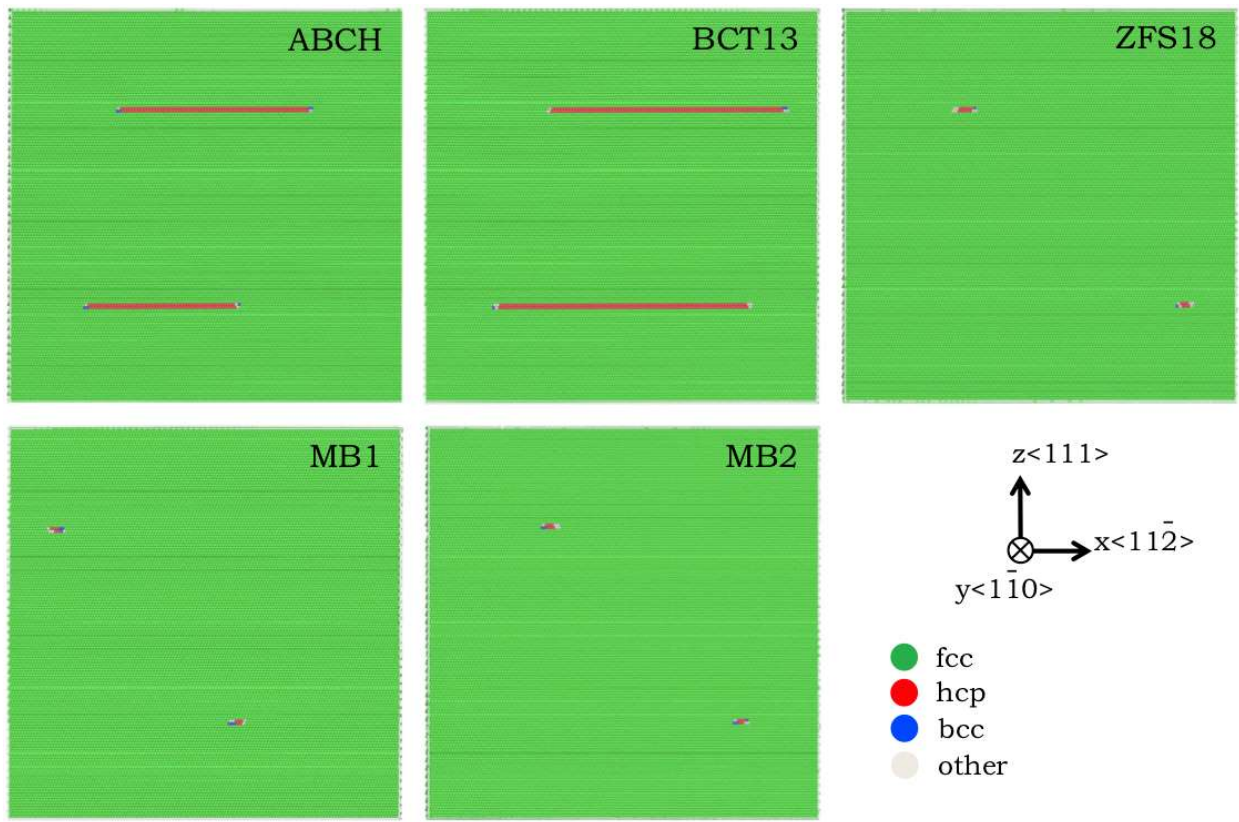

Figure 10. Snapshots of the simulation cells with screw dislocation dipoles at $T$=1273 K. The ABCH97 and BCT13 potentials lead to very large separations between partials because these potentials provide very low SFEs. The ZFS18, MB1 and MB2 potential lead to much smaller separations between partials.

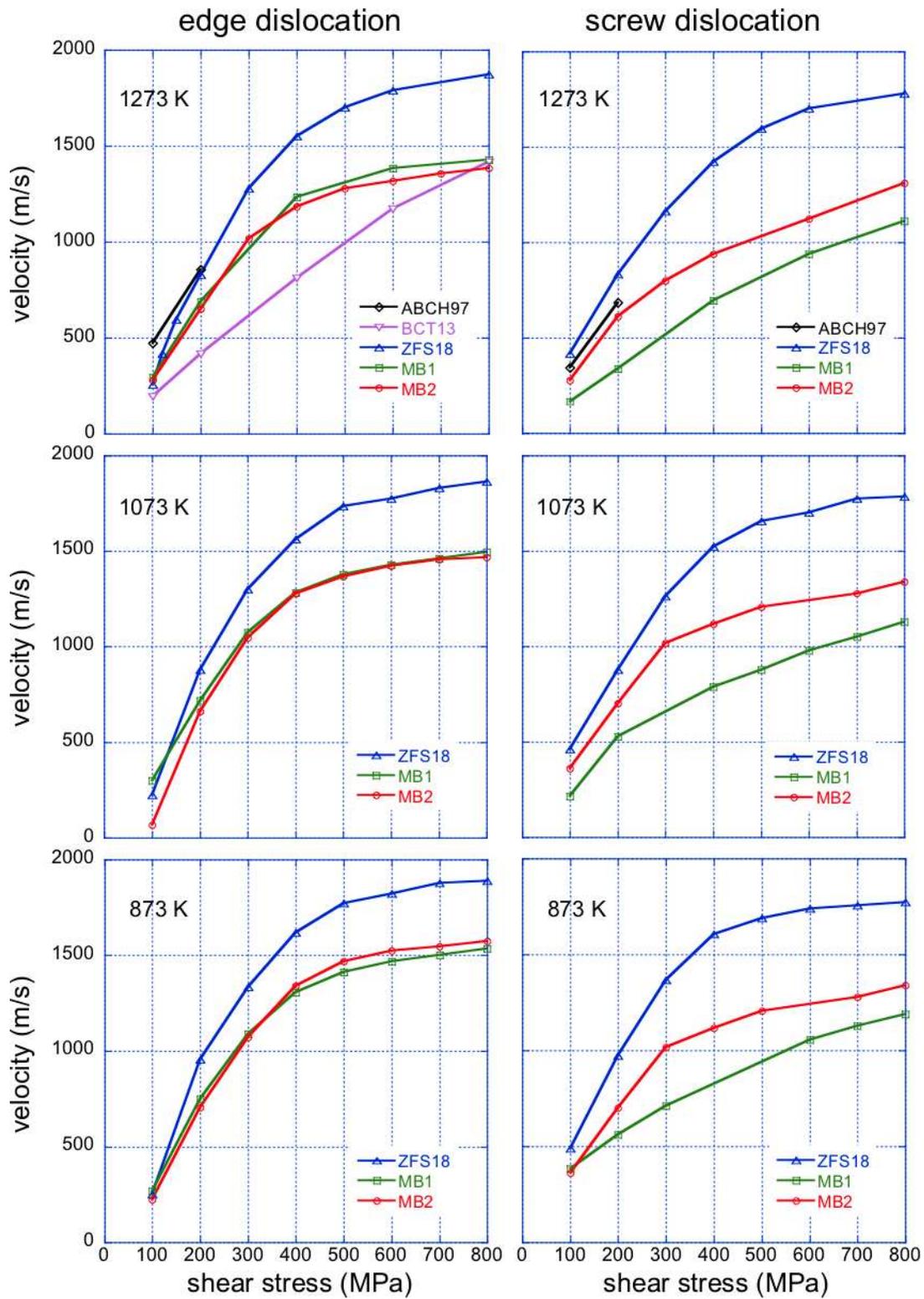

Figure 11. Dislocation velocities as functions of the applied stress.

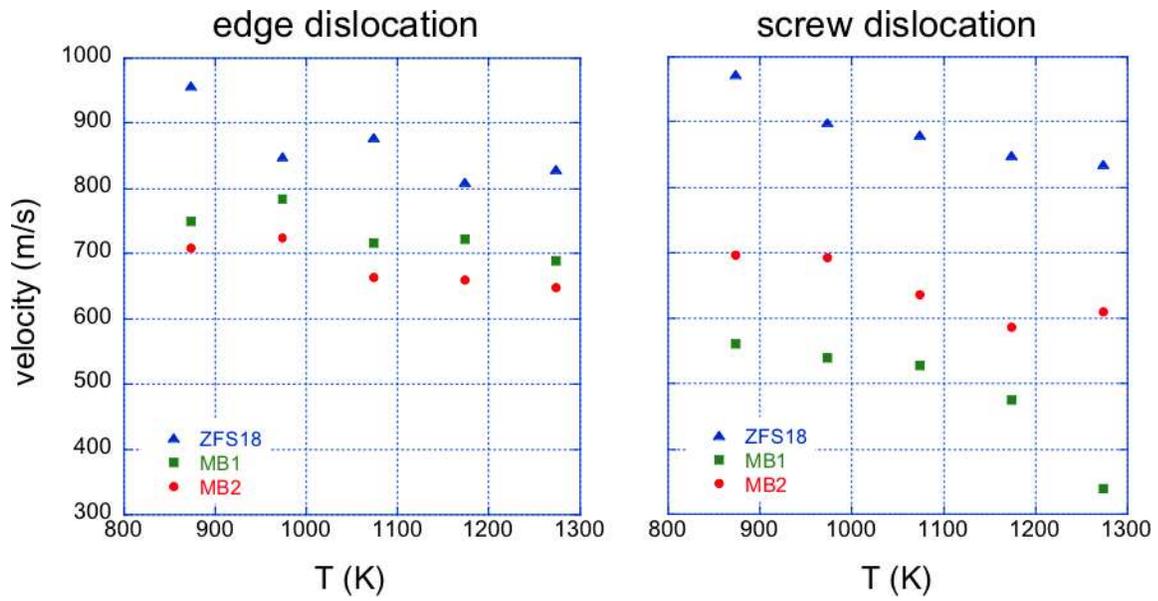

Figure 12. Temperature dependences of the dislocation velocities for the shear stress σ=0.2 GPa.

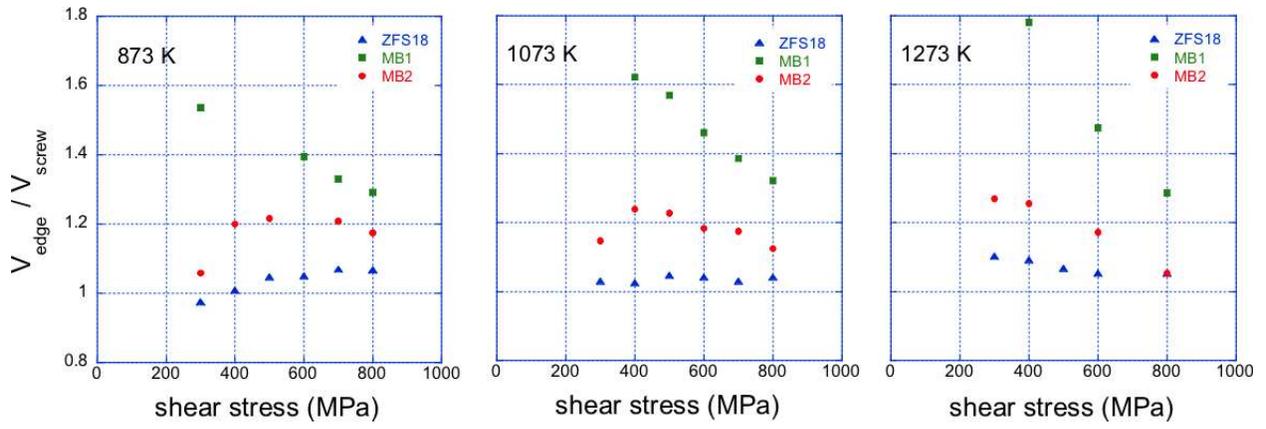

Figure 13. The ratios of the edge and screw dislocation velocities as the functions of applied shear stress.